\documentclass[aps,twocolumn,showpacs]{revtex4}

\usepackage{amsfonts,epsfig}

\newcommand{\HH}{\mathcal{H}}
\newcommand{\PP}{\mathcal{P}}
\def\ua{\uparrow}
\def\da{\downarrow}

\begin{document}

\title{Quantum Walk on a Line with Two Entangled Particles}

\author{Y. Omar}
\affiliation{GoLP, Centro de F\'isica de Plasmas, Instituto
Superior T\'ecnico, P-1049-001 Lisbon, Portugal}

\author{N. Paunkovi\'c, L. Sheridan} \affiliation{Centre for
Quantum Computation, Clarendon Laboratory, University of Oxford,
Parks Road, Oxford OX1 3PU, United Kingdom}

\author{S. Bose}
\affiliation{Department of Physics and Astronomy, University
College London, Gower Street, London WC1E 6BT, United Kingdom}

\date{9 November 2004}


\begin{abstract}
We introduce the concept of a quantum walk with two particles and
study it for the case of a discrete time walk on a line. A quantum
walk with more than one particle may contain entanglement, thus
offering a resource unavailable in the classical scenario and
which can present interesting advantages. In this work, we show
how the entanglement and the relative phase between the states
describing the \textit{coin} degree of freedom of each particle
will influence the evolution of the quantum walk. In particular,
the probability to find at least one particle in a certain
position after $N$ steps of the walk, as well as the average
distance between the two particles, can be larger or smaller than
the case of two unentangled particles, depending on the initial
conditions we choose. This resource can then be tuned according to
our needs, in particular to enhance a given application
(algorithmic or other) based on a quantum walk. Experimental
implementations are briefly discussed.

\end{abstract}


\maketitle

\begin{figure}[ht]
\begin{center}
\epsfig{file=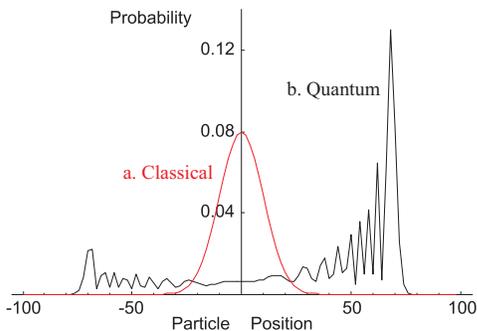, width=2.5in}
\end{center}
\caption{Probability distribution for a classical random walk (a)
on a line after $N=100$ steps, as well as for a quantum walk (b)
with initial state $|0 \rangle \otimes |\!\! \ua \rangle$ and a
Hadamard coin.} \label{Fig. One-particle}
\end{figure}

Quantum walks, the quantum version of random walks, were first
introduced in 1993 \cite{Aharonov} and have since then been a
topic of research within the context of quantum information and
computation (for an introduction, see \cite{Kempe}). Given the
superposition principle of Quantum Mechanics, quantum walks allow
for coherent superpositions of classical random walks and, due to
interference effects, can exhibit different features and offer
advantages when compared to the classical case. In particular, for
a quantum walk on a line, the variance after $N$ steps is
proportional to $N$, rather than $\sqrt{N}$ as in the classical
case (see Fig.\ \ref{Fig. One-particle}).  Recently, several
quantum algorithms with optimal efficiency were proposed based on
quantum walks \cite{Shenvi}, and it was even shown that a
continuous time quantum walk on a specific graph can be used for
exponential algorithmic speed-up \cite{Childs}.

All studies on quantum walks so far have, however, been based on a
single walker. In this article we study a discrete time quantum
walk on a line with two particles. Classically, random walks with
$K$ particles are equivalent to $K$ independent single-particle
random walks. In the quantum case though, a walk with $K$
particles may contain entanglement, thus offering a resource
unavailable in the classical scenario which can present
interesting advantages. Moreover, in the case of identical
particles we have to take into account the effects of quantum
statistics, giving an additional feature to quantum walks that can
also be exploited. In this work we explicitly show that a quantum
walk with two particles can indeed be tuned to behave very
differently from two independent single-particle quantum walks.
This paves the way for new quantum algorithms based on richer
quantum walks.

Let us start by introducing the discrete time quantum walk on a
line for a single particle. The relevant degrees of freedom are
the particle's position $i$ (with $i \! \in \! \mathbb{Z}$) on the
line, as well as its \emph{coin} state. The total Hilbert space is
given by $\HH \equiv \HH_P \otimes \HH_C$, where $\HH_P$ is
spanned by the orthonormal vectors $\{|i\rangle\}$ representing
the position of the particle, and $\HH_C$ is the two-dimensional
coin space spanned by two orthonormal vectors which we denote as
$|\!\! \ua \rangle$ and $|\!\! \da\rangle$.

Each step of the quantum walk is given by two subsequent
operations. First, the \emph{coin operation}, given by $\hat{U}_C
\! \in \! SU(2)$ and acting only on $\HH_C$, which will be the
quantum equivalent of randomly choosing which way the particle
will move (like tossing a coin in the classical case). The
non-classical character of the quantum walk is precisely here, as
this operation allows for superpositions of different
alternatives, leading to different moves. Then, the
\emph{shift-position operation} $\hat{S}$ will move the particle
accordingly, transferring this way the quantum superposition to
the total state in $\HH$. The evolution of the system at each step
of the walk can then be described by the total unitary operator:
\begin{equation}
\label{Eq. U} \hat{U} \equiv \hat{S} (\hat{I}_P \otimes
\hat{U}_C),
\end{equation}
where $\hat{I}_P$ is the identity operator on $\HH_P$. Note that
if a measurement is performed after each step, we will revert to
the classical random walk.

In this article we choose to study a quantum walk with a
\emph{Hadamard coin}, i.e.\ where $\hat{U}_C$ is the Hadarmard
operator $\hat{H}$:
\begin{equation}
\hat{H}=\frac{1}{\sqrt{2}} \left[
\begin{array}{cc}
1 & 1 \\ 1 & -1
\end{array}
\right].
\end{equation}
Note that this represents a \emph{balanced coin}, i.e.\ there is a
fifty-fifty chance for each alternative. The shift-position
operator is given by:
\begin{equation}
\hat{S} =  \left( \sum_i |i+1 \rangle \langle i | \right) \otimes
|\!\! \ua \rangle \langle \ua \!\!| + \left( \sum_i |i-1\rangle
\langle i | \right) \otimes | \!\! \da \rangle \langle \da \!\! |.
\end{equation}
Therefore, if the initial state of our particle is, for instance
$|0 \rangle \otimes |\!\! \ua \rangle$, the first step of the
quantum walk will be as follows:
\begin{eqnarray}
|0 \rangle \otimes |\!\! \ua \rangle
&\stackrel{\hat{H}}{\longrightarrow}& |0\rangle \otimes
\frac{1}{\sqrt{2}} \left( |\!\! \ua\rangle + |\!\! \da\rangle
\right) \nonumber \\
&\stackrel{\hat{S}}{\longrightarrow}& \frac{1}{\sqrt{2}} \left(
|1\rangle \otimes |\!\! \ua\rangle + |-1\rangle \otimes |\!\!
\da\rangle \right).
\end{eqnarray}
We see that there is a probability of $1/2$ to find the particle
in position $1$, as well as to find it in position $2$, just like
in the classical case. Yet, if we let this quantum walk evolve
beyond the (two) initial steps before we perform a position
measurement, we will find a very different probability
distribution for the position of the particle when compared to the
classical random walk, as it can be seen in Fig.\ \ref{Fig.
One-particle} for $N=100$ steps.


Let us consider the previous quantum walk, but now with two
non-interacting particles on a line (not necessarily the same). If
the particles are distinguishable and in a separable state, the
position measurement of one particle will not change the
probability distribution of the other, they are completely
uncorrelated. On the other hand, if the particles are entangled, a
new resource with no classical equivalent will be at our disposal.

The joint Hilbert space of our composite system is given by:
\begin{equation}
\HH_{12} \equiv \HH_{1} \otimes \HH_{2} \equiv (\HH_{P,1} \otimes
\HH_{C,1}) \otimes (\HH_{P,2} \otimes \HH_{C,2}),
\end{equation}
where $\HH_{1}$ and $\HH_{2}$ represent the Hilbert spaces of
particles $1$ and $2$ respectively. Since the relevant degrees of
freedom in our problem are the same for both particles, we have
that both $\HH_{1}$ and $\HH_{2}$ are isomorphic to $\HH$ defined
earlier for the one-particle case. Note also that in the case of
identical particles we have to restrict $\HH_{12}$ to its
symmetrical and antisymmetrical subspaces, respectively for bosons
and fermions.

Let us then consider the case where both particles start the
quantum walk in the same position, $0$, but with different coin
states $|\!\! \da\rangle$ and $|\!\! \ua\rangle$. In the case
where the particles are in a pure separable state, our system's
initial state will be given by:
\begin{equation}
\label{Eq. Initial-S}
|\psi_0^S\rangle_{12}=|0,\da\rangle_1|0,\ua\rangle_2.
\end{equation}
We can also consider an initial pure state entangled in the coin
degrees of freedom. In particular, we will consider the following
two maximally entangled states:
\begin{equation}
\label{Eq. Initial-E}
|\psi_0^{\pm}\rangle_{12}=\frac{1}{\sqrt{2}}(|0,\da\rangle_1|0,\ua\rangle_2
\pm |0,\ua\rangle_1|0,\da\rangle_2),
\end{equation}
differing only by a relative phase. Note that if we were
considering identical particles on the same point, our system
would have to be described by these states, for bosons and
fermions respectively.

\begin{figure}[ht]
\begin{center}
\epsfig{file=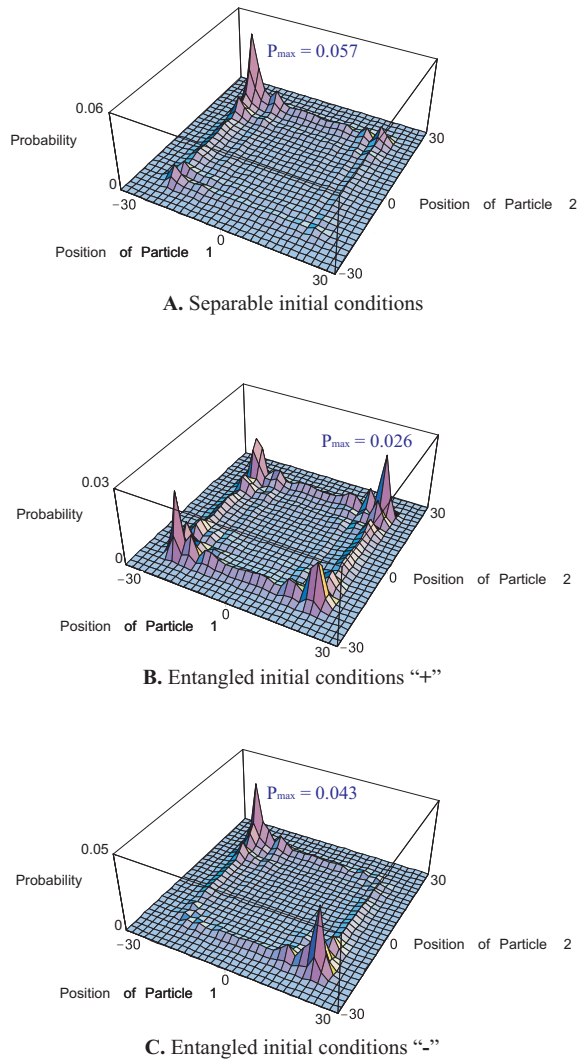, width=3in}
\end{center}
\caption{Two-particles probability distributions after $N=30$
steps for different initial conditions: (A) separable state
$|\psi_0^S\rangle_{12}$; (B) $|\psi_0^+\rangle_{12}$ state; and
(C) $|\psi_0^-\rangle_{12}$ state. Note the different vertical
ranges.} \label{Fig. Plots}
\end{figure}

Each step of this two-particles quantum walk will be given by:
\begin{equation}
\hat{U}_{12}= \hat{U} \otimes \hat{U},
\end{equation}
where $\hat{U}$ is given by equation (\ref{Eq. U}) and is the same
for both particles. After $N$ steps, the state of the system will
be, in the case of the initial conditions (\ref{Eq. Initial-S}):
\begin{equation}
|\psi_N^S\rangle_{12}=\hat{U}^N_{12} \, |\psi_0^S\rangle_{12}=
\hat{U}^N |0,\da\rangle_1 \hat{U}^N |0,\ua\rangle_2.
\end{equation}
Fig.\ \ref{Fig. Plots}.a shows the joint probability distribution
$P_{12}^S (i,j;N)$ for finding particle $1$ in position $i$ and
particle $2$ in position $j$ for $N=30$ steps. Note that, since
the particles are uncorrelated, $P_{12}^S (i,j;N)$ is simply the
product of the two independent one-particle distributions:
\begin{equation}
P_{12}^S (i,j;N)=P^S_1(i;N) \times P^S_2(j;N),
\end{equation}
where $P^S_1(i;N)$ is the probability distribution for finding
particle $1$ in position $i$ after $N$ steps, and similarly for
$P^S_2(j;N)$ and particle $2$. This can also be observed in Fig.\
\ref{Fig. Plots}.a, which is clearly the product of  two
distributions like the one in Fig.\ \ref{Fig. One-particle}.b, one
biased to the left for particle $1$ and the other to the right for
particle $2$, accordingly with the initial conditions given by
equation (\ref{Eq. Initial-S}).

In the case of entangled particles, the state of the system after
$N$ steps will be:
\begin{eqnarray}
& & |\psi_N^{\pm}\rangle_{12} = \hat{U}^N_{12} \,
|\psi_0^{\pm}\rangle_{12}
\\ & = & \frac{1}{\sqrt{2}}\left( \hat{U}^N |0,\da\rangle_1 \, \hat{U}^N |0,\ua\rangle_2 \pm
\hat{U}^N |0,\ua\rangle_1 \, \hat{U}^N |0,\da\rangle_2 \right)
\nonumber.
\end{eqnarray}
The probability distribution for finding particle $1$ in position
$i$ and particle $2$ in position $j$ in the ``+" case, $P_{12}^+
(i,j;N)$, is represented in Fig.\ \ref{Fig. Plots}.b, for $N=30$.
Similarly, Fig.\ \ref{Fig. Plots}.c shows the distribution
$P_{12}^- (i,j;N)$ of the ``-" case, again for $N=30$. The effects
of the entanglement are striking when comparing the three
distributions in Fig.\ \ref{Fig. Plots}. In particular, we see
that these effects significantly increase the probability of the
particles reaching certain configurations on the line, which
otherwise would be very unlikely to be occupied. In all cases, the
maxima of the distributions occur around positions $\pm\, 20$. In
the ``+" case it is most likely to find both particles together,
whereas for ``-" the former situation is impossible, and the
particles will tend to finish as distant as possible from one
another.

Let us now consider the individual particles in the entangled
system. The state of each of them can be described by the reduced
density operator $\hat{\rho}_{1,2}(N) \equiv Tr_{2,1} \left(
|\psi_N^{\pm}\rangle_{12} \,\, _{12}\langle \psi_N^{\pm}|
\right)$, which consists of an equal mixture of the one-particle
states $\hat{U}^N|0,\da\rangle$ and $\hat{U}^N|0,\ua\rangle$.
Thus, given one particle, the probability to find it in position
$i$ after $N$ steps is given by the following marginal probability
distribution:
\begin{equation}
P_{1,2}^{\pm}(i;N) = \frac{1}{2} \left[ P_{\da}(i;N) +
P_{\ua}(i;N) \right].
\end{equation}
where $P_{\da}(i;N)$ is the probability distribution for finding
the particle in state $\hat{U}^N|0,\da\rangle$ in position $i$
after $N$ steps, and similarly for $P_{\ua}(i;N)$ and a particle
in state $\hat{U}^N|0,\ua\rangle$. Note that in the case of
initial conditions (\ref{Eq. Initial-S}) we have the marginal
probabilities $P^S_1(i;N) = P_{\da}(i;N)$ and $P^S_2(i;N) =
P_{\ua}(i;N)$. But now, contrary to the separable case, the joint
probability $P_{12}^{\pm}(i,j;N)$ is no longer the simple product
of the two one-particle probabilities, as it contains information
about the non-trivial correlations between the outcomes of the
position measurement performed on each particle. Therefore, to
investigate more quantitatively the difference between quantum
walks with two distinguishable particles in a pure separable state
and two entangled particles in their coin degree of freedom, we
must look at joint (two-particle) rather than individual
properties.


First, let us consider the distance between the two particles,
$\Delta^{S,\pm}_{12}$. For that, we define the random variables
$x_1$ and $x_2$ as the outcomes of the (single particle)
measurement of the position of particle $1$ and particle $2$
respectively, after $N$ steps of the quantum walk (they can take
integer values between $N$ and $-N$). We can now define the
distance as, for the three different initial conditions:
\begin{equation}
\Delta^{S,\pm}_{12} \equiv |x_1 - x_2|.
\end{equation}
The expectation value $\langle \Delta^{S,\pm}_{12} \rangle$ of the
distance is given in Table \ref{Tab. Distance} for different $N$.
We see that in the ``-" case the particles tend, on average, to
end the quantum walk more distant to each other, whereas in the
``+" case they tend to stay closer, and somewhere in between in
the separable case. In fact, for fixed $N$, we always have:
\begin{equation}
\langle\Delta^{-}_{12}\rangle - \langle\Delta^{S}_{12}\rangle =
\langle\Delta^{S}_{12}\rangle - \langle\Delta^{+}_{12}\rangle.
\end{equation}

Let us now consider the correlation function between the spatial
distribution of each of the two particles:
\begin{equation}
C^{S,\pm}(x_1,x_2) \equiv \langle x_1 x_2 \rangle - \langle x_1
\rangle\langle x_2 \rangle.
\end{equation}
Clearly, in the case of the separable state (\ref{Eq. Initial-S})
this correlation is always zero. For maximally entangled
particles, the values of $C^{\pm}(x_1,x_2)$ are presented in Table
\ref{Tab. Correlation} for different $N$. Given the symmetry of
the $P_{12}^{\pm}$ distributions, we have that in those cases
$\langle x_1 \rangle = \langle x_2 \rangle = 0$. Thus, the sign
difference in the correlation function expresses the tendency for
the two particles in the ``-" case to end the quantum walk on
different sides of the line (with respect to its origin 0), and on
the same side for the ``+" case.

Finally, let us now calculate, for the different initial
conditions, the probability of finding at least one particle in
position $i$ after $N$ steps: $\PP^{S,\pm}(i;N)$. This is clearly
a joint property as it depends on both one-particle outcomes:
\begin{eqnarray}
& & \PP^{S,\pm}(i;N) = \nonumber
\\ & = & \sum_{j=-N}^N \left[P_{12}^{S,\pm}(i,j;N) + P_{12}^{S,\pm}(j,i;N)\right]
- P_{12}^{S,\pm}(i,i;N) \nonumber
\\ & = & \left[P_{1}^{S,\pm}(i;N)+P_{2}^{S,\pm}(i;N)\right] -
P_{12}^{S,\pm}(i,i;N) \nonumber
\\ & = & \left[P_{\da}(i;N)+P_{\ua}(i;N)\right] - P_{12}^{S,\pm}(i,i;N).
\end{eqnarray}
Given a one-particle probability distribution, say $P_{\da}(i;N)$
[note that $P_{\ua}(i;N)=P_{\da}(-i;N)$], our probability
$\PP^{S,\pm}(i;N)$ decreases with the joint probability
$P_{12}^{S,\pm}(i,i;N)$ and is maximal in the ``-" case, as we
always have $P_{12}^-(i,i;N)=0$. In fact, around the points
$(20,20)$ and $(-20,-20)$ in Fig. \ref{Fig. Plots} we clearly
have:
\begin{equation}
\PP^-(i;N)>\PP^S(i;N)>\PP^+(i;N).
\end{equation}
We see that, by introducing entanglement in the initial conditions
of our two-particles quantum walk, the probability of finding at
least one particle in a particular position on the line can
actually be better or worst than in the case where the two
particles are independent. Note that in this case this does not
depend on the particular amount of entanglement introduced, as
both states in Eq. (\ref{Eq. Initial-E}) are maximally entangled,
but rather on their symmetry/relative phase.
\\


\begin{table}
\begin{tabular}{*{16}{c}}
\multicolumn{16}{c}{\textbf{Expectation value $\langle\Delta^{S,\pm}_{12} \rangle$ after $N$ steps}}\\[0.5mm]
\hline & Nb. of steps $N$ & & & 10 & & 20 & & 30 & & 40 & & 60 & & 100 & \\
\hline & Init. cond. $|\psi_0^-\rangle_{12}$ & & & 8.8 & & 17.5 & & 26.0 & & 34.9 & & 52.2 & & 87.0 & \\
\hline & Init. cond. $|\psi_0^S\rangle_{12}$ & & & 7.1 & & 14.7 & & 21.9 & & 29.5 & & 44.3 & & 73.9 & \\
\hline & Init. cond. $|\psi_0^+\rangle_{12}$ & & & 5.5 & & 11.9 & & 17.8 & & 24.1 & & 36.3 & & 60.8 & \\
\hline
\end{tabular}
\caption{Average distance $\langle \Delta^{S,\pm}_{12} \rangle$
after $N$ steps.}
\label{Tab. Distance}
\end{table}

\begin{table}
\begin{tabular}{*{14}{c}}
\multicolumn{14}{c}{\textbf{Correlation function $C^{S, \pm}(x_1,x_2)$ after $N$ steps}}\\[0.5mm]
\hline & Nb. of steps $N$  &  10 & & 20 & & 30 & & 40 & & 60 & & 100 & \\
\hline & Init. c. $|\psi_0^-\rangle_{12}$ &  -16.8 & & -69.8 & & -153.5 & & -276.2 & & -619.7 & & -1718.3 & \\
\hline & Init. c. $|\psi_0^S\rangle_{12}$ &  0 & & 0 & & 0 & & 0 & & 0 & & 0 & \\
\hline & Init. c. $|\psi_0^+\rangle_{12}$ &  4.8 & & 7.3 & & 13.7 & & 15.1 & & 23.1 & & 39.1 & \\
\hline
\end{tabular}
\caption{Correlation function $C^{S, \pm}(x_1,x_2)$ after $N$
steps.}
\label{Tab. Correlation}
\end{table}


This work opens way for generalization in a number of ways. First,
one could consider periodic or other boundary conditions on the
line, or more general graphs \cite{Dorit}. Note that the positions
on the line of our two particles could also be interpreted as the
position of a single particle doing the quantum walk on a regular
two-dimensional lattice. More general coins could also be
considered \cite{Hillery,Oli}, including entangling and
non-balanced coins, as well as different initial states. One could
also augment the number of particles, study these quantum walks in
continuous time or in their asymptotic limit. Furthermore, also
very interesting and promising is to investigate the use of
multiparticle quantum walks in the design of quantum algorithms
\cite{Mateus}, or in solving mathematical or practical problems
that could be encoded has a quantum walk, such as the estimation
of the volume of a convex body \cite{Convex} or the connectivity
in a P2P network \cite{P2P}.

Finally, some brief comments about implementations of our
two-particles quantum walk on a line. The methods recently
proposed for the single-particle case using cavity QED
\cite{Knight}, optical lattices \cite{Kendon} or ion traps
\cite{Travaglione} could be adapted to our two-particles case. For
instance, in the latter we could encode the coin states in the
electronic levels of two ions and the position in their
\textit{com} or \textit{stretch} motional modes: the coin flipping
could then be obtained with a $\pi/2$ Raman pulse and the shift
with a conditional optical dipole force \cite{Schaetz}. Another
possibility is to send two photons through a tree of balanced beam
splitters which implement both the coin flipping and the
conditional shift, again generalizing a scheme proposed for a
single particle \cite{Hillery,Kim}. Note that this could be
implemented with other particles as well, e.g.\ electrons, using a
device equivalent to a beam splitter \cite{Yamamoto}. Also very
interesting, regardless of any particular method or technology, is
the possibility of using two indistinguishable particles on the
same line to implement our quantum walk. Say we encode the coin
degrees of freedom in the polarization of two photons, or in the
spin of two electrons: if the two particles start in the same
position $0$, then they will be forced to be in state given by Eq.
(\ref{Eq. Initial-E}) (``+" for bosons and ``-" for fermions).
Although the particles will initially be only entangled in the
mathematical sense (as they cannot be addressed to extract quantum
correlations), this is a perfectly valid way of preparing the
$|\psi_0^+\rangle_{12}$ and $|\psi_0^-\rangle_{12}$ initial states
for our quantum walk, saving us the trouble of generating
entangled pairs. Thus, the indistinguishability of identical
particles appears once again as a resource for quantum information
processing \cite{Omar}, here in particular offering a way to
simplify the preparation of the initial states for a two-particle
quantum walk on a line.

In this article we introduced the concept of a quantum walk with
two particles and studied it for the case of a discrete time walk
on a line. Having more than one particle, we could now add a new
feature to the walk: entanglement between the particles. In
particular, we considered initial states maximally entangled in
the coin degrees of freedom and with opposite symmetries, and
compared them to the case where the two particles were initially
unentangled and thus independent. We found that the entanglement
in the coin states introduced spatial correlations between the
particles, and that their average distance is larger in the ``-"
case than in the separable case, and is smaller in the ``+" case.
This could benefit algorithmic applications which require two
marked sites to be reached which are known {\em a priori} to be on
the opposite or on the same sides of a line. We also found that
the introduction of entanglement could increase or decrease the
probability to find at least one particle on a given point of the
line. This increase could allow us to reach a marked site faster
than with two unentangled quantum walkers. The entanglement in the
initial conditions thus appears as a resource that we can tune
according to our needs to enhance a given application (algorithmic
or other) based on a quantum walk.


\begin{acknowledgments}

The authors would like to thank K. Banaszek, E. Kashefi, P. Mateus
and T. Schaetz for precious discussions and help. YO acknowledges
support from Funda\c{c}\~{a}o para a Ci\^{e}ncia e a Tecnologia
(Portugal) and the 3rd Community Support Framework of the European
Social Fund, the QuantLog initiative, and wishes to thank the
Clarendon Laboratory for their hospitality. NP thanks Elsag S.p.A.
for financial support. LS, currently at the Institute for Quantum
Computing in Waterloo, gratefully acknowledges the support of the
Nuffield Foundation through their Undergraduate Research Bursary.
Finally, YO and SB would like to thank the Institute for Quantum
Information at Caltech, where this work was started, for their
hospitality.

\end{acknowledgments}



\begin{thebibliography}{99}

%
\bibitem{Aharonov} Y. Aharonov, L. Davidovich, and N. Zagury, Phys. Rev. A \textbf{48}, 1687
(1993).
%
\bibitem{Kempe} J. Kempe, Contemp. Phys. \textbf{44}, 307 (2003).
%
\bibitem{Shenvi} N. Shenvi, J. Kempe, and K. B. Whaley, Phys. Rev. A \textbf{67}, 052307
(2003); A. Ambainis, quant-ph/0311001 (2003).
%
\bibitem{Childs} A. M. Childs \emph{et al.}, \emph{Proc. 35th ACM Symposium on Theory of
Computing (STOC}), 59 (2003).
%
\bibitem{Dorit} D. Aharonov, A. Ambainis, J. Kempe, and U. Vazirani,
\textit{Proceedings 33rd ACM Symposium on Theory of Computation
(STOC)}, 50 (2001); T.D. Mackay, S.D. Bartlett, L.T. Stephenson,
and B.C. Sanders, J. Phys. A: Math. Gen. \textbf{35}, 2745 (2002);
B. Tregenna, W. Flanagan, R. Maile, and V. Kendon, New J. Phys.
{\bf 5}, 83 (2003).
%
\bibitem{Hillery} M. Hillery, J. Bergou, and E. Feldman, Phys. Rev. A \textbf{68}, 032314 (2003).
%
\bibitem{Oli} O. Buerschaper and K. Burnett, quant-ph/0406039 (2004).
%
\bibitem{Mateus} P. Mateus and Y. Omar, work in progress.
%
\bibitem{Convex} M. Dyer, A. Frieze, and R. Kannan, Journal of the ACM \textbf{38}, 1 (1991).
%
\bibitem{P2P} Q. Lv \textit{et al.}, \textit{Proceedings of the 16th ACM
International Conference on Supercomputing (ICS)}, 84 (2002).
%
\bibitem{Knight} B. C. Sanders, S. D. Bartlett, B. Tregenna, and P. L. Knight,
Phys. Rev. A \textbf{67}, 042305 (2003).
%
\bibitem{Kendon} W. D\"ur, R. Raussendorf, V. M. Kendon, and H.-J.
Briegel, Phys. Rev. A \textbf{66}, 052319 (2002).
%
\bibitem{Travaglione} B. C. Travaglione and G. J. Milburn, Phys.
Rev. A \textbf{65}, 032310 (2002).
%
\bibitem{Schaetz} T. Schaetz, work in progress.
%
\bibitem{Kim} H. Jeong, M. Paternostro, and M. S. Kim, Phys. Rev. A \textbf{69}, 012310 (2004).
%
\bibitem{Yamamoto} R. C. Liu, B. Odom, Y. Yamamoto, and S. Tarucha, Nature {\bf 391}, 263 (1998).
%
\bibitem{Omar} Y. Omar, N. Paunkovi\'c, S. Bose, and V. Vedral, Phys. Rev. A \textbf{65}, 062305
(2002); N. Paunkovi\'c, Y. Omar, S. Bose and V. Vedral, Phys. Rev.
Lett. \textbf{88}, 187903 (2002).
%

\end{thebibliography}
\end{document}